# Unsupervised Deep Contrast Enhancement with Power Constraint for OLED Displays

Yong-Goo Shin, Seung Park, Yoon-Jae Yeo, Min-Jae Yoo and Sung-Jea Ko, *Fellow*, IEEE
School of Electrical Engineering, Korea University, Seoul 136-713, Republic of Korea
E-mail: sjko@korea.ac.kr

*Abstract*—Various power-constrained contrast enhancement (PCCE) techniques have been applied to an organic light emitting diode (OLED) display for reducing the power demands of the display while preserving the image quality. In this paper, we propose a new deep learning-based PCCE scheme that constrains the power consumption of the OLED displays while enhancing the contrast of the displayed image. In the proposed method, the power consumption is constrained by simply reducing the brightness a certain ratio, whereas the perceived visual quality is preserved as much as possible by enhancing the contrast of the image using a convolutional neural network (CNN). Furthermore, our CNN can learn the PCCE technique without a reference image by unsupervised learning. Experimental results show that the proposed method is superior to conventional ones in terms of image quality assessment metrics such as a visual saliency-induced index (VSI) and a measure of enhancement (EME).[1]

## I. Introduction

ORGANIC light emitting diode (OLED) displays have widely replaced liquid crystal displays (LCDs), owing to several factors such as high brightness, fine viewing angle, and the possibility of building thin and flexible screens [1]. Even though OLED displays have high power efficiency, they place a significant strain on the battery, *e.g.*, in mobile devices and home television sets. Therefore, various techniques such as hardware optimization [2] and image processing-based methods [3]-[11] have been proposed to reduce the power demands of the OLED displays.

As an OLED directly emits light at each pixel, the total power consumption of the display is known to be proportional to the sum of the power consumption of each pixel [3]-[5].

However, as the brightness of the overall pixels is reduced to save power, the visibility and contrast of the image with reduced brightness are also decreased. To cope with this problem, Lee *et al.* [3] proposed a histogram-based power-constrained contrast enhancement (HPCCE) technique that saves power while enhancing the contrast of the image to preserve visual quality. A variant of HPCCE based on the multiscale retinex was introduced in [4], whereas Jang *et al.* [5] proposed a non-iterative HPCCE method for fast computation. Peng *et al.* [6] combined a histogram shrinking technique with contrast enhancement to achieve the same goal as HPCCE for real-time applications of full-HD displays. However, as the aforementioned HPCCE-based methods do not set a constraint on the maximum image modification, they sometimes produce an unpredictable resultant image [7]. To cope with this problem, Pagliari *et al.* [7] recently proposed an adaptive power saving and contrast enhancement technique called LAPSE that simultaneously constrains the power consumption and maximizes the contrast of the image, while limiting the amount of image modification.

Some researchers, [8], [10], and [11], proposed a power-saving scheme that limits image modification based on image quality assessment (IQA) metrics, such as the structural similarity metric (SSIM) [9]. Chang *et al.* [8] proposed an SSIM-based pixel dimming method with a constraint on the maximum image modification, and this method is further improved by correcting the overexposed region as introduced in [10]. Recently, Chondro *et al.* [11] proposed a hue-preserving pixel dimming technique to avoid dynamic region distortions.

In this paper, we propose a novel deep learning method for power-constrained contrast enhancement (PCCE). However, supervised learning cannot be employed for PCCE since there is no standard reference image of the PCCE algorithm. Moreover, if the output image of a traditional PCCE method is used as an alternative of the reference image, the trained network cannot generate a higher quality image than the reference image. To solve this problem, this paper presents an unsupervised learning framework for PCCE. In the proposed method, the power consumption is constrained by simply reducing the brightness by a certain ratio, whereas the perceived visual quality is preserved as much as possible by enhancing the contrast of the image using the convolutional neural network (CNN). For the proposed unsupervised PCCE method, we design a loss function that can simultaneously optimize for OLED power loss, similarity to a given input image, and degree of contrast enhancement. We conduct extensive experiments to demonstrate that our method outperforms conven-

[1]This research was funded and supported by LG Display CO., Ltd.

Y.-G. Shin is with School of Electrical Engineering Department, Korea University, Anam-dong, Sungbuk-gu, Seoul, 136-713, Rep. of Korea (e-mail: ygshin@dali.korea.ac.kr).

S. Park is with School of Electrical Engineering Department, Korea University, Anam-dong, Sungbuk-gu, Seoul, 136-713, Rep. of Korea (e-mail: spark@dali.korea.ac.kr).

Y.-J. Yeo is with School of Electrical Engineering Department, Korea University, Anam-dong, Sungbuk-gu, Seoul, 136-713, Rep. of Korea (e-mail: yjyeo@dali.korea.ac.kr).

M.-J. Yoo is with LG Display, Magok-dong, Gangseo-gu, Seoul, 07796, Rep. of Korea (e-mail : mjyoo0125@lgdisplay.com).

S.-J. Ko is with School of Electrical Engineering Department, Korea University, Anam-dong, Sungbuk-gu, Seoul, 136-713, Rep. of Korea (e-mail: sjko@korea.ac.kr).

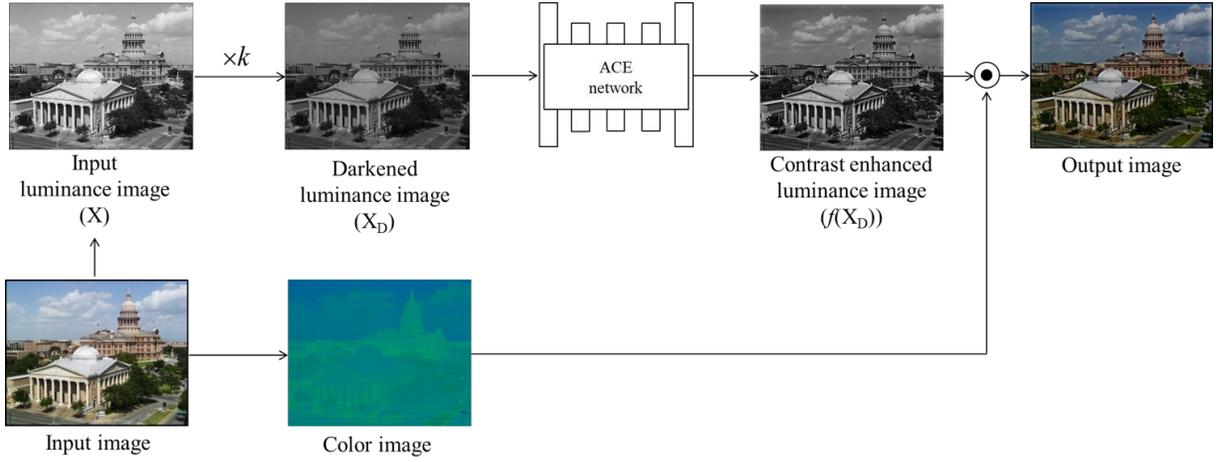

Fig. 1. The framework of the proposed method. In the proposed method, the luminance image X is first scaled down to reduce the power consumption. Then, the darkened luminance image $X_D$ goes through the CNN to generate the contrast-enhanced image $f(X_D)$. By combining the color image and the contrast-enhanced luminance image, we finally obtain the output image.

tional methods on popular datasets, such as the Berkeley segmentation dataset (BSD) [34] and LIVE [35]. Experimental results show that the proposed method is superior to the conventional methods in terms of IQA metrics: visual saliency-induced index (VSI) [12], and a measure of enhancement (EME) [36]. In summary, the major contributions of this paper are as follows:

(*i*) This paper is the first work that successfully introduces unsupervised learning to PCCE.

(*ii*) We propose a novel loss function which guides the networks to learn PCCE without the reference image.

## II. PRELIMINARIES

### A. Power Model for OLED Displays

The total dissipated power (TDP) of an OLED display with $N$ pixels is given by [3], [4], and [7]:

$$P = \sum_{i=1}^{N} \left( w_o + w_R R_i^\gamma + w_G G_i^\gamma + w_B B_i^\gamma \right), \quad (1)$$

where $R_i$, $G_i$, and $B_i$ are the red, green, and blue intensities of the $i$-th pixel, respectively, $w_R$, $w_G$, $w_B$ and $\gamma$ are panel-dependent coefficients, and $w_o$ is a constant value representing the supplementary power required for the non-pixel part of the display. Note that $w_o$ is ignored, *i.e.* $w_o = 0$, in most literature such as [3], [4], and [7]. In [4], the TDP is approximated by using only the Y-component in the YUV color space, *i.e.* luminance, as follows:

$$P = \sum_{i=1}^{N} Y_i^\gamma, \quad (2)$$

where $Y_i$ is the intensity of the Y-component of the $i$-th pixel. In the remainder of this paper, we use Eq. (2) to compute the TDP of the image.

### B. Deep Learning-Based Image Enhancement

The CNN has been widely adopted to not only computer vision studies, *e.g.* semantic segmentation [13]-[15], but also low-level vision problems such as image dehazing or image enhancement scheme [16]-[20]. In particular, deep learning-based image enhancement techniques have shown superior performance compared to non-deep learning methods as demonstrated in [17]-[20]. These methods often employed (or modified) the popular CNN architectures such as U-Net [21] and VGG network [22] as their base networks and introduced the novel loss functions which encourage the network to effectively conduct the target enhancement algorithm. Even though these methods exhibit fine performance, they require high hardware costs owing to numerous network parameters; these networks are difficult for the hardware implementation.

Recently, Chen *et al.* [23] and Talebi *et al.* [24] applied a compact network called a multiscale context aggregation network (CAN) [25] to the image enhancement. More specifically, the CAN consists of multiple dilated convolutional layers with increasing dilation rates to aggregate the global contextual information. Since the CAN requires a small number of network parameters compared with U-Net or VGG network, it is more suitable to implement on the hardware; the CAN is more appropriate for applying to PCCE than U-Net or VGG network. Therefore, we employ the CAN as the base network of the proposed method.

However, the drawback of CAN-based methods [23], [24] is that their performance cannot exceed that of an existing algorithm since they train their networks by using the output of the conventional method as the reference image. In other words, when applying the CAN-based methods to PCCE scheme without any modification, it could not show the better performance compared to the conventional method which is used for producing the reference image. To solve this problem, this paper introduces a novel loss function for unsupervised learning-based PCCE method.

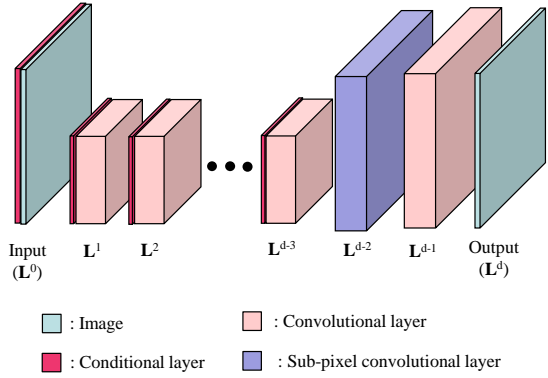

Input (**L**[0])    **L**[1]    **L**[2]    **L**[d-3]    **L**[d-2]    **L**[d-1]    Output (**L**[d])

☐ : Image    ☐ : Convolutional layer
☐ : Conditional layer    ☐ : Sub-pixel convolutional layer

Fig. 2. The architecture of the ACE network. The ACE network effectively enhances the contrast while preserving the power saving rate of the luminance-reduced image.

## III. PROPOSED METHOD

In general, training a deep CNN model for a specific task involves two components: (*i*) network architecture design and (*ii*) loss function for training the network. For network architecture design, as mentioned in Section II.B, we employ the modified CAN as the base network owing to its advantage that requiring a small number of network parameters. For loss function, we jointly optimize three loss terms to train the network without the reference image, *i.e.* unsupervised learning. Fig. 1 illustrates the framework of the proposed method. In the proposed method, the input image is decomposed into luminance and color images. Then, the intensity of the luminance image is reduced by a certain ratio for reducing the power consumption. The contrast of the luminance-reduced (or darkened) image is enhanced by using the CNN and then combined with the decomposed color image. In the next subsection, we explain the adaptive contrast enhancement (ACE) network for the PCCE.

### A. ACE Network

The darkened luminance image $X_D$ with power saving rate $R$, $R = 1 - (P_{dim}/P_{in})$ where $P_{in}$ and $P_{dim}$ are the TDP of input and dimmed images, is generated by reducing the intensity of the luminance image X by a certain ratio $k$. The relationship between $R$ and $k$ is readily obtained by using Eq. (2) as follows:

$$R = 1 - \frac{P_{dim}}{P_{in}} = 1 - \frac{\sum_i^N (kY_i)^\gamma}{\sum_i^N (Y_i)^\gamma} = 1 - k^\gamma. \quad (3)$$

With respect to $k$, the Eq. (3) can be rewritten by $k = (1-R)^{\frac{1}{\gamma}}$. However, pure scaling of the luminance can significantly degrade the perceived image quality. To cope with this problem, we introduce an ACE network that enhances the contrast of the dimmed image. Fig. 2 shows an architecture of the ACE network, consisting of multiple consecutive layers: {$\mathbf{L}^0$, ..., $\mathbf{L}^d$}, where the first and last layers $\mathbf{L}^0$ and $\mathbf{L}^d$ represent the input and output images, respectively. In order to alleviate the number of convolution operations, we first project $\mathbf{L}^0$ into $\mathbf{L}^1$ by performing $2 \times 2$ average pooling after a convolutional layer with 32 feature maps and kernels of $3 \times 3$ size. Then, similar to [23]-[25], we use the CAN architecture with 32 feature maps and kernels of $3 \times 3$ size for intermediate layers $\mathbf{L}^i$, $1 \le i \le d-3$. In the intermediate layers, to aggregate the global contextual information, we employ the multiple dilated convolutional layers with different dilation rates. In the ACE network, the dilation rate of $\mathbf{L}^i$ is set to $2^i$. By enlarging the receptive field, the ACE network is able to reflect the characteristic of the whole image for enhancing the contrast.

To produce $\mathbf{L}^{d-2}$ having the same resolution with the input image, we utilize a sub-pixel convolutional layer [26], which up-scales the low-resolution feature maps to high-resolution feature maps. The output image, *i.e.* contrast enhanced luminance image $f(X_D)$, is obtained through the convolutional layer with 32 feature maps and $3 \times 3$ kernels, followed by linear transformation ($1 \times 1$ convolutional with no nonlinearity). We use the leaky rectified linear unit (LReLU) as an activation function in all convolutional layers, except $\mathbf{L}^d$. To avoid artifacts generated by zero padded layers with high dilation rates, we use reflection padding for all layers.

To adaptively control a contrast enhancement degree according to the given $R$ value, the ACE network has to produce different feature maps depending on the given $R$ value. In other words, the intermediate layers need additional information about the $R$ value. To this end, in this paper, we adopt the concept of a conditional generative adversarial network (cGAN) [27], which generates the resultant image according to a given condition. In general, cGAN obtains the information about the given condition by concatenating a conditional label (*e.g.*, binary or one-hot vector) to the input or intermediate layers. In the ACE network, we first make a layer with single channel having $R$ value, which is denoted as *Conditional layer* in Fig. 2. Then, this layer is concatenated as the conditional label to the input and multiple different intermediate layers. This approach encourages the input and intermediate layers to adaptively produce different feature maps depending on the given $R$ value, which allows the ACE network to control the degree of contrast enhancement. In the next subsection, we introduce the loss function for the network training procedure.

### B. Loss Function

To train the ACE network, we jointly optimize three different loss terms: *power loss* $L_p$ ensures that the power saving rate $R$ is maintained after contrast enhancement, *similarity loss* $L_s$ limits the amount of image alteration, and *contrast loss* $L_c$ drives the network to maximize the contrast of the output image. The total loss function is defined as a weighted sum over these three loss terms:

$$L_{total} = \lambda_p L_p + \lambda_s L_s + \lambda_c L_c, \quad (4)$$

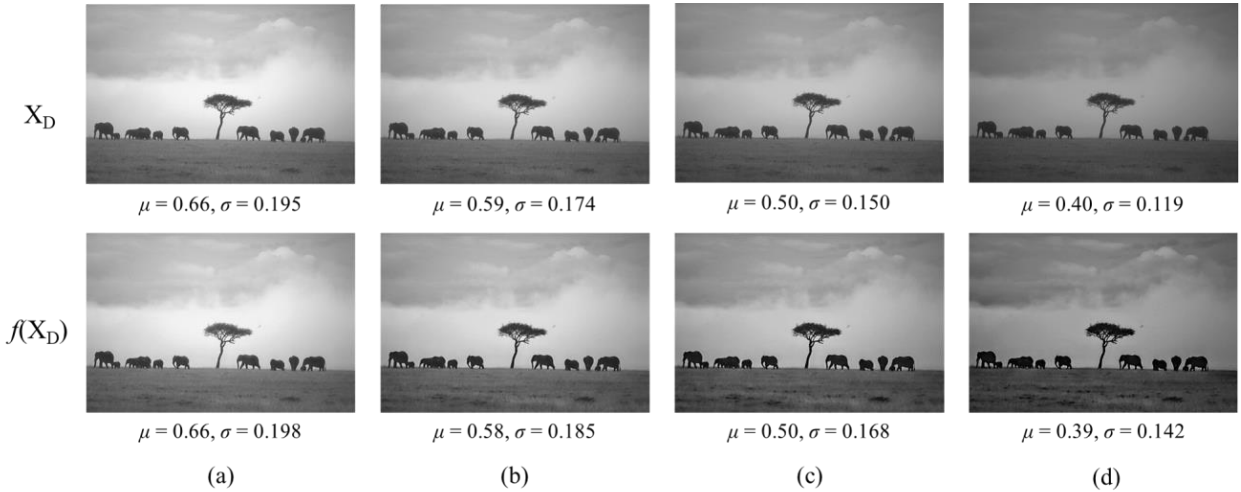

Fig. 3. The variations of the $\mu$ and $\sigma$ according to the $R$ value. (a), (b), (c), and (d) illustrate the darkened luminance images $X_D$ and the contrast enhanced ones $f(X_D)$ when $R$ value is 0.1, 0.3, 0.5, and 0.7, respectively.

where $\lambda_p$, $\lambda_s$ and $\lambda_c$ are hyper-parameters that control the relative importance of each term. Details on each loss term will be provided in the rest of this subsection.

**Power loss** A goal of the ACE network is to not only enhance the contrast but also maintain the $R$ value after contrast enhancement. Thus, we define $L_P$ as the simple absolute difference between the $R$ values of $X_D$ and $f(X_D)$, i.e., $R$ and $\hat{R}$, which is expressed as $L_P = |R - \hat{R}|$.

**Similarity loss** Most deep learning-based image processing techniques train their network using $l_1$ or $l_2$ loss functions that measure a difference between output and reference images. However, as mentioned in Section II.B, our approach does not have a reference image, which is necessary to employ $l_1$ or $l_2$ loss functions. Instead of these loss functions, and motivated by [28], we use an SSIM loss that drives the network to preserve structural characteristics of the input luminance image. The SSIM loss is defined as follows:

$$L_s = 1 - \text{SSIM}(X, f(X_D)). \quad (5)$$

**Contrast loss** Pagliari et al. [7] demonstrated that the global contrast can be effectively enhanced by increasing the standard deviation of the luminance image. However, the drawbacks of this approach are that details of input image are lost and the local contrast is decreased. In order to enhance the global and local contrasts simultaneously, $L_C$ is defined as a combination of global and local contrast losses,

$$L_c = \lambda_G L_c^G + L_c^L, \quad (6)$$

where $L_c^G$ and $L_c^L$ are the global and local contrast loss terms, respectively, and $\lambda_G$ is a hyper-parameter controlling the contributions from each loss term. More specifically, $L_c^G$ and $L_c^L$

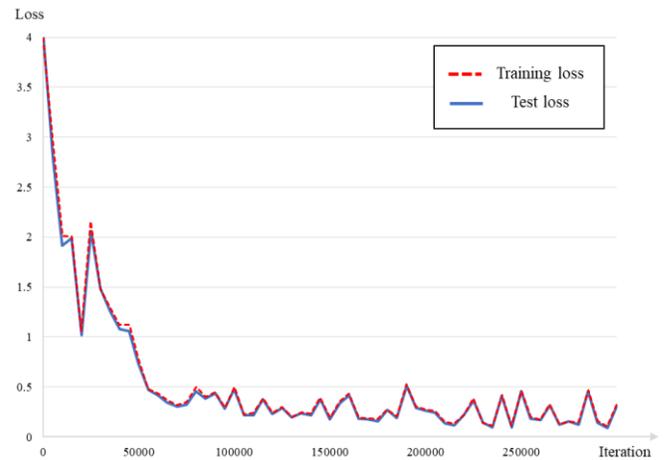

Fig. 4. The loss graph during the training procedure. The red dashed line indicate the training loss, whereas the blue line is the test loss.

mainly focus on maximizing the standard deviation of luminance while preserving the mean brightness in the whole image and the local patch, respectively. To this end, we define $L_c^G$ as follows:

$$L_c^G = \left|\mu_{X_D} - \mu_{f(X_D)}\right| - (1 - w_{std})\ln\left(\frac{\sigma_{f(X_D)}}{\sigma_{X_D}}\right) + w_{std}\left|\sigma_{X_D} - \sigma_{f(X_D)}\right|, \quad (7)$$

$$w_{std} = \max(0, 1 - 2R), \quad (8)$$

where $\mu_{X_D}$ ($\mu_{f(X_D)}$) and $\sigma_{X_D}$ ($\sigma_{f(X_D)}$) represent the mean and variance of the $X_D$ ($f(X_D)$), respectively. In the same manner, we obtain $L_c^L$ by averaging local contrast losses of whole pixels. More specifically, the local contrast loss of each pixel, $L_{c,i}^L$, is computed by using a local patch with $11 \times 11$ size, which is expressed as follows:

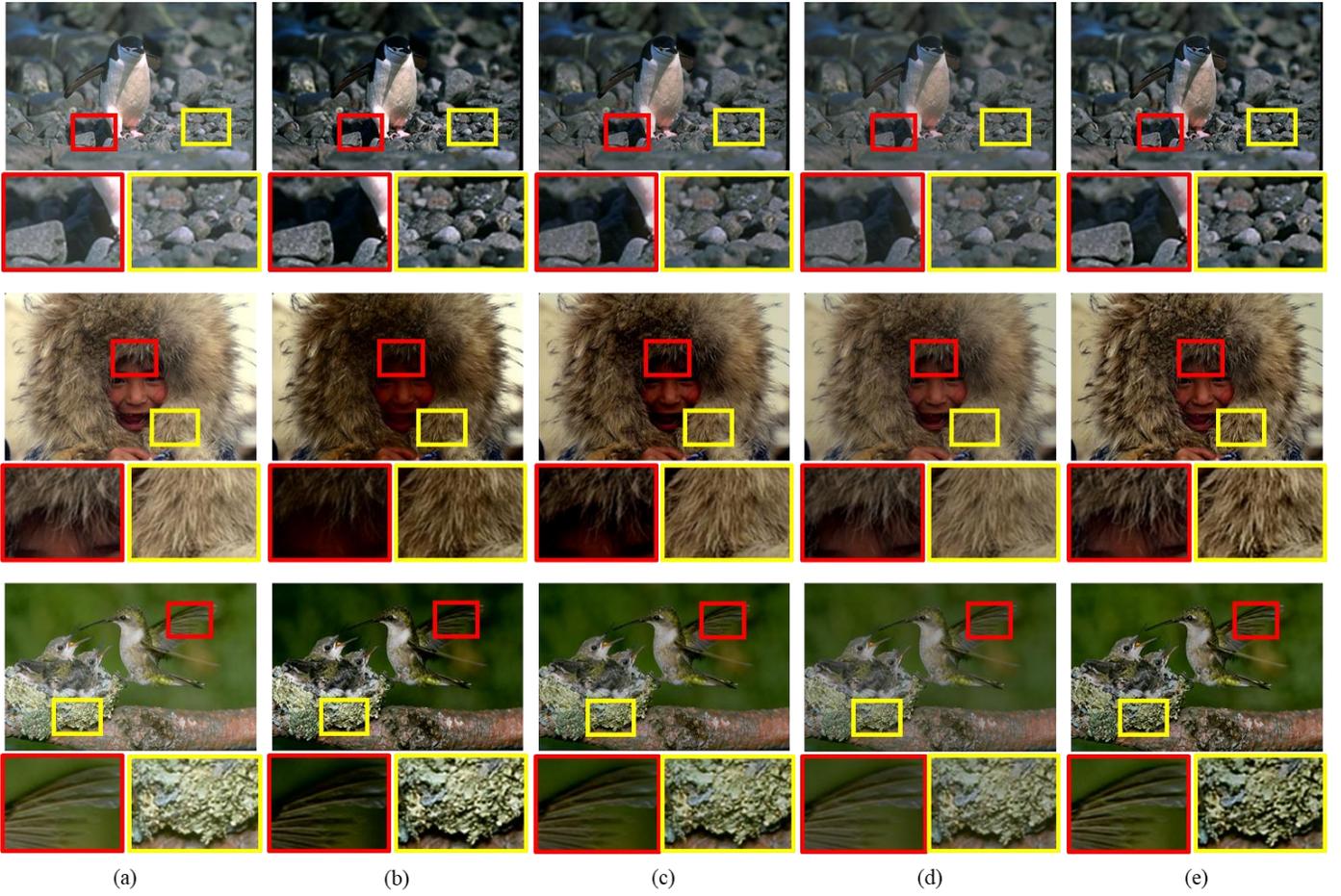

Fig. 5. The visual comparison of the proposed method and conventional methods in BSD test sets. The resultant images for $R = 0.5$. (a) Input image, (b) HPCCE [3], (c) Chang *et al.* [8], (d) Chondro *et al.* [11], (e) proposed method. The proposed method produces a resultant image with superior image quality as compared to conventional methods.

$$L^L_{c,i} = \left|\mu_{P_i(X_D)} - \mu_{P_i(f(X_D))}\right| - (1 - w_{std})\ln\left(\frac{\sigma_{P_i(f(X_D))}}{\sigma_{P_i(X_D)}}\right)$$
$$+ w_{std}\left|\sigma_{P_i(X_D)} - \sigma_{P_i(f(X_D))}\right|, \quad (9)$$

where $\mu_{P_i(X_D)}$ ($\mu_{P_i(f(X_D))}$) and $\sigma_{P_i(X_D)}$ ($\sigma_{P_i(f(X_D))}$) indicate the mean and variance of $11 \times 11$ patch at $i$-th pixel in $X_D$ ($f(X_D)$), respectively. Note that $L^G_c$ and $L^L_c$ drive the ACE network to adaptively generate the output image according to the $R$ value. For smaller $R$, $L^G_c$ and $L^L_c$ encourage the ACE network to maintain $\mu$ and $\sigma$. In contrast, a larger $R$ results in a contrast-enhanced image with a larger $\sigma$ while retaining $\mu$. Fig. 3 shows $\mu$ and $\sigma$ of the $X_D$ and $f(X_D)$ with various $R$ values, *i.e.* $R$=0.1, 0.3, 0.5, and 0.7. As shown in Fig. 3, the network preserves the $\mu$ value regardless of the $R$ value, whereas the $\sigma$ is increased as the $R$ value becomes larger.

### C. Implementation Details

The ACE network is trained for 300k iterations with a learning rate of 0.0001 in an end-to-end manner. In this paper, we set a batch size as 1, which is widely used in the various deep learning techniques [29]-[32]. As all parameters in the ACE network are differentiable, we performed optimization employing the Adam optimizer [33], which is a stochastic optimization method with adaptive estimation of moments. We set the parameters of Adam optimizer $\beta_1$ and $\beta_2$ to 0.9 and 0.999, respectively. The ACE network consists of seven layers, *i.e.* $d$=7, and the hyper-parameters were empirically set to ($\lambda_P$, $\lambda_s$, $\lambda_c$, $\lambda_G$) = (10, 2, 0.25, 2) to control the relative importance of each loss term. In particular, to achieve the target $R$ while maintaining the structure information, we provided more constraints on both $L_p$ and $L_s$ compared with $L_c$. For each iteration during the training process, the $R$ was randomly selected in the range of [0.01, 0.8] for each image. In the test process, we use a certain $R$ in a range from 0.1 to 0.7 with 0.1 intervals.

Our experiments were conducted on a single Titan X GPU (Pascal architecture) and implemented in *Tensorflow*. For our experiments, we use the BSD dataset [34] including 500 images to randomly sample 400 images as a training set and 100 images as a test set. The original resolution of the image is utilized to train the ACE network. Fig. 4 shows the loss graph during the training procedure. We recorded the average $L_{total}$ values of the training and test datasets, *i.e.* training loss and test loss, respectively Note that we employed the random $R$ value when the training and test losses are measured. As depicted in Fig. 4, the test loss follows the training loss, which indicates that the ACE network is trained stably without resulting in the over-fitting problem.

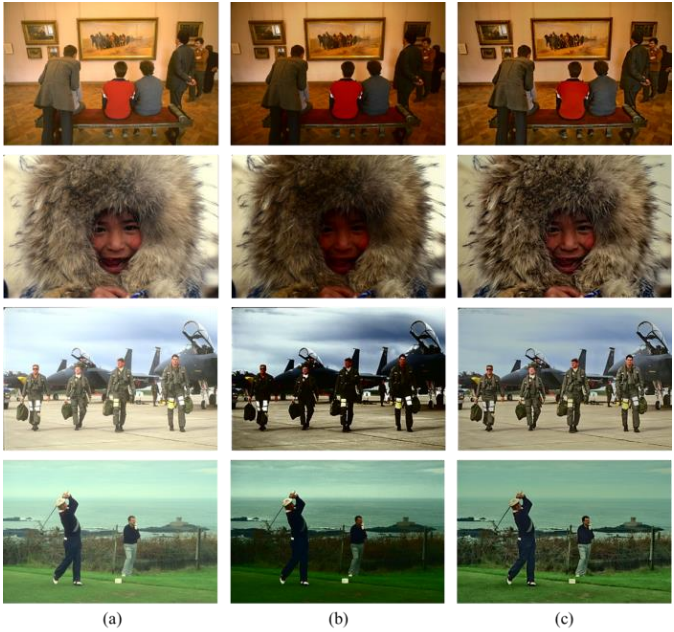

Fig. 6. The visual comparison of LAPSE and the proposed method in BSD test sets. For fair comparison, we determine the R value of ACE network such that the LAPSE and the proposed method saves the same amount of the power. (a) Input image, (b) LAPSE [7], (c) proposed method.

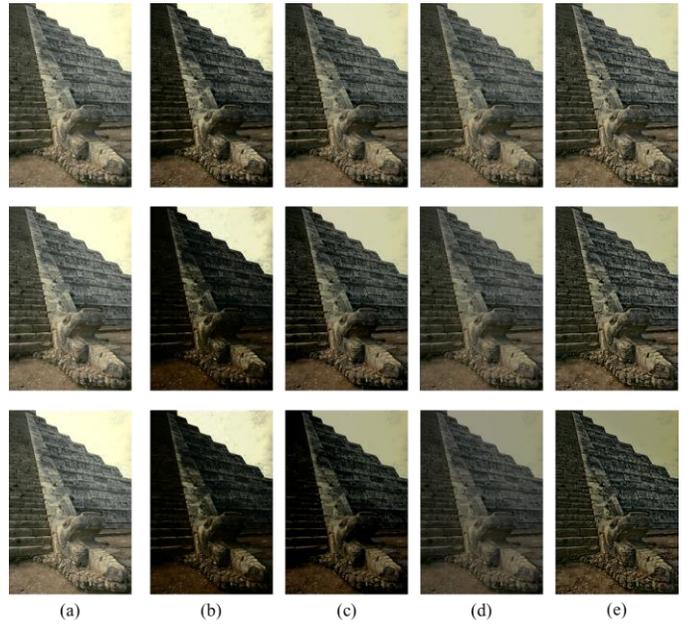

Fig. 7. The visual comparison of the proposed method and conventional ones. Each row represents the output image for R=0.3, R=0.5, and R=0.7, respectively. (a) Input image without saving power, (b) HPCCE [3], (c) Chang *et al.* [8], (d) Chondro *et al.* [11], (e) proposed method.

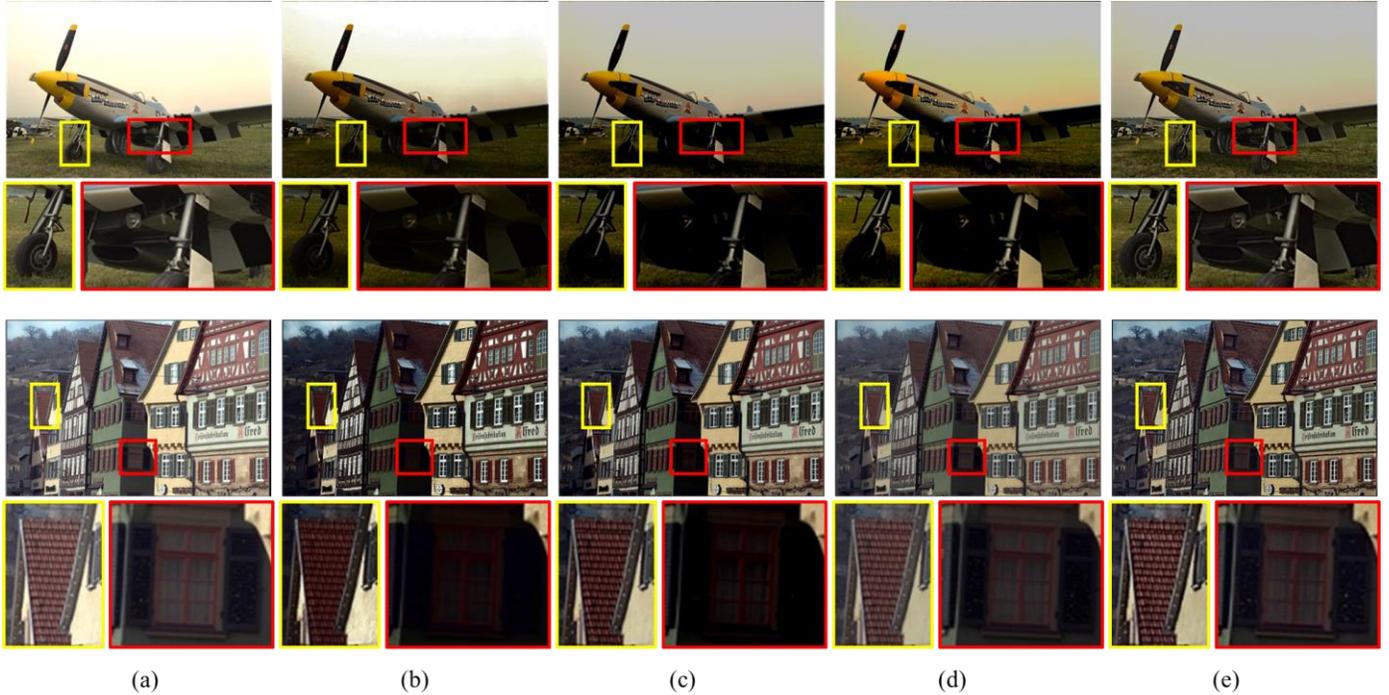

Fig. 8. The visual comparison of the proposed method and conventional ones in LIVE datasets. The resultant images are produced for *R*=0.5. (a) Input image, (b) HPCCE [3], (c) Chang *et al.* [8], (d) Chondro *et al.* [11], (e) proposed method.

## IV. EXPERIMENTAL RESULTS

In this section, we show qualitative and quantitative comparison results to demonstrate the superiority of the proposed method. In this study, we compare the proposed method with state-of-the-art power-constrained image enhancement methods including [3], [7], [8], and [11]. However, as the conven-tional methods do not allow for fixing the desired $R$, it is difficult to analyze the overall performance for equal power consumption. To solve this problem, we empirically select hyperparameters to produce an output image with the desired $R$ for the conventional methods. All of our experiments are conducted by changing the $R$ value from 0.1 to 0.7 with 0.1 intervals.

TABLE I
OBJECTIVE PERFORMANCE COMPARISON OF THE CONVENTIONAL METHODS WITH THE PROPOSED METHOD

| Data set | $R$ | EME | | | | NDE | | | | VSI | | | | $R_a$ |
|---|---|---|---|---|---|---|---|---|---|---|---|---|---|---|
| | | HPCCE [3] | Chang et al. [8] | Chondro et al. [11] | ACE network (Ours) | HPCCE [3] | Chang et al. [8] | Chondro et al. [11] | ACE network (Ours) | HPCCE [3] | Chang et al. [8] | Chondro et al. [11] | ACE network (Ours) | |
| BSD | 0.1 | 10.63 | 7.77 | 7.84 | 7.75 | 0.602 | 0.493 | 0.493 | 0.493 | 0.990 | 1.000 | 0.999 | 0.996 | 0.100 |
| | 0.2 | 11.00 | 8.83 | 7.83 | 8.60 | 0.579 | 0.476 | 0.462 | 0.479 | 0.989 | 0.999 | 0.999 | 0.995 | 0.200 |
| | 0.3 | 11.39 | 10.02 | 7.82 | 9.63 | 0.543 | 0.452 | 0.432 | 0.464 | 0.985 | 0.998 | 0.998 | 0.994 | 0.300 |
| | 0.4 | 11.92 | 11.45 | 7.80 | 10.80 | 0.499 | 0.422 | 0.404 | 0.447 | 0.979 | 0.995 | 0.997 | 0.992 | 0.400 |
| | 0.5 | 12.41 | 13.04 | 7.78 | 12.05 | 0.450 | 0.383 | 0.376 | 0.425 | 0.971 | 0.990 | 0.994 | 0.989 | 0.501 |
| | 0.6 | 12.89 | 14.66 | 7.74 | 13.24 | 0.398 | 0.337 | 0.347 | 0.397 | 0.960 | 0.982 | 0.989 | 0.986 | 0.601 |
| | 0.7 | 13.25 | 16.06 | 7.71 | 14.21 | 0.346 | 0.288 | 0.317 | 0.363 | 0.946 | 0.969 | 0.981 | 0.982 | 0.701 |
| | **Avg** | **11.93** | **11.69** | **7.79** | **10.90** | **0.488** | **0.407** | **0.404** | **0.438** | **0.974** | **0.990** | **0.994** | **0.991** | **-** |
| LIVE | 0.1 | 9.68 | 8.56 | 8.91 | 7.98 | 0.593 | 0.493 | 0.488 | 0.498 | 0.996 | 1.000 | 0.998 | 0.998 | 0.098 |
| | 0.2 | 10.12 | 9.71 | 8.89 | 8.89 | 0.557 | 0.474 | 0.455 | 0.486 | 0.995 | 0.999 | 0.998 | 0.997 | 0.200 |
| | 0.3 | 10.57 | 11.07 | 8.86 | 9.98 | 0.536 | 0.450 | 0.424 | 0.470 | 0.992 | 0.998 | 0.997 | 0.996 | 0.300 |
| | 0.4 | 11.08 | 12.47 | 8.82 | 11.21 | 0.492 | 0.422 | 0.397 | 0.450 | 0.987 | 0.995 | 0.996 | 0.994 | 0.400 |
| | 0.5 | 11.66 | 13.81 | 8.78 | 12.50 | 0.442 | 0.391 | 0.368 | 0.425 | 0.979 | 0.991 | 0.993 | 0.991 | 0.501 |
| | 0.6 | 12.29 | 15.39 | 8.72 | 13.73 | 0.385 | 0.349 | 0.338 | 0.394 | 0.965 | 0.983 | 0.988 | 0.988 | 0.601 |
| | 0.7 | 12.79 | 16.69 | 8.68 | 14.75 | 0.331 | 0.299 | 0.307 | 0.357 | 0.948 | 0.970 | 0.979 | 0.983 | 0.701 |
| | **Avg** | **11.17** | **12.53** | **8.80** | **11.29** | **0.477** | **0.411** | **0.397** | **0.440** | **0.980** | **0.991** | **0.993** | **0.992** | **-** |

TABLE II
OBJECTIVE PERFORMANCE COMPARISON OF LAPSE WITH THE PROPOSED METHOD

| Data set | | LAPSE [7] | Proposed method |
|---|---|---|---|
| BSD | EME | 10.478 | 11.735 |
| | NDE | 0.373 | 0.431 |
| | VSI | 0.979 | 0.989 |
| LIVE | EME | 10.691 | 12.425 |
| | NDE | 0.365 | 0.426 |
| | VSI | 0.934 | 0.955 |

TABLE III
EFFECTIVENESS OF CONTRAST LOSS ON BSD

| | $R$ | (i) | (ii) | (iii) | (iv) |
|---|---|---|---|---|---|
| EME | 0.1 | 6.778 | 7.065 | 7.517 | 7.75 |
| | 0.3 | 7.204 | 7.814 | 8.946 | 9.63 |
| | 0.5 | 7.732 | 8.755 | 10.667 | 12.05 |
| | 0.7 | 8.357 | 9.799 | 12.372 | 14.21 |
| | **Avg** | **7.518** | **8.358** | **9.876** | **10.90** |

### A. Qualitative Comparison

We show qualitative comparison results for $R = 0.5$ in Fig. 5. Fig. 5(b) and Fig. 5(c) show the results of the HPCCE method [3] and the method in [8], respectively. Note that the resultant images have a high global contrast, but the image details are not well preserved. For instance, most details of cloth in the first images are lost. As shown in Fig. 5(d), the method of Chondro et al. [11] preserves details relatively well, but produces a low contrast image with degraded image quality. More specifically, as depicted in the yellow box of Fig. 5(d), the method in [11] cannot effectively enhance the image contrast. Fig. 5(e) shows the resultant images of the proposed method, where not only are the image details properly preserved, but also the global contrast is enhanced.

In the LAPSE method, it is difficult to obtain an output image with the desired $R$ because the power is minimized under the mean SSIM (MSSIM) constraint [9]. For this reason, we first produce the resultant image of the LAPSE method with MSSIM = 0.75. Then, we adjust the $R$ value for the proposed ACE network, such that the LAPSE and the proposed method save the same amount of power. Fig. 6 shows the visual comparison of LAPSE and the proposed method. As depicted in Fig. 6, the LAPSE method exhibits the detail loss due to the over-enhancement, whereas the proposed method not only preserves details but also efficiently enhances the contrast without introducing visual artifacts.

To perform qualitative evaluation under various $R$ values, as shown in Fig. 7, we also generate the output images of the proposed method and conventional methods. The proposed method generates an output image with better quality as compared to conventional ones, for all $R$ values. In particular, when $R$ value becomes large, the methods in [3] and [8] lost details. In addition, as depicted in the third row of Fig. 7(d), the method in [11] results in low image contrast. In contrast, the proposed method is able to enhance the contrast while effectively preserving the details. Thus, these results indicate

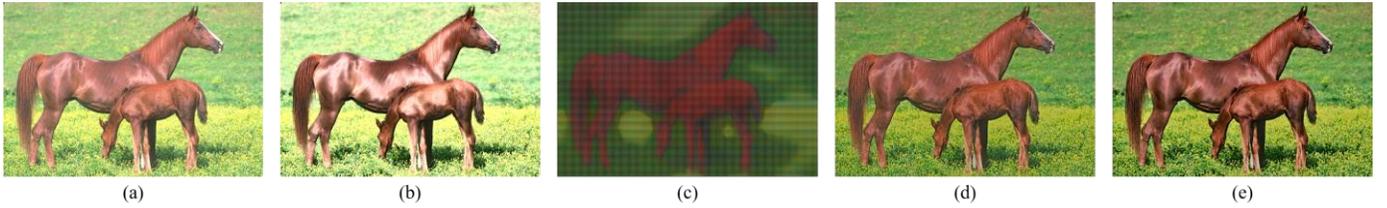

Fig. 9. The visual comparisons of the proposed method trained without each loss term, *i.e.* $L_p$, $L_s$, and $L_c$. In our experiments, we set $R$ to 0.5. (a) Input image, (b) Resultant image of model trained without $L_p$, (c) Resultant image of model trained without $L_s$, (d) Resultant image of model trained without $L_c$, (e) Resultant image of proposed method.

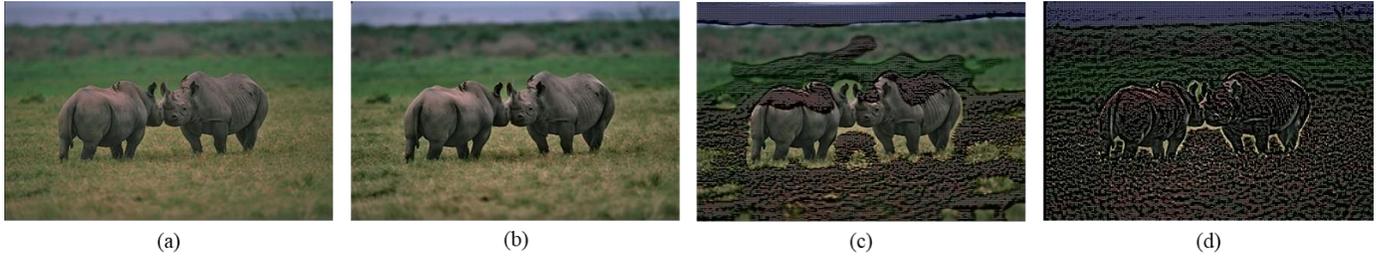

Fig. 10. The visual comparisons of the proposed method trained by using the different $\lambda_c$ values. In our experiments, we set $R$ to 0.5. (a) $\lambda_c = 0.1$, (b) $\lambda_c = 0.25$, (c) $\lambda_c = 0.5$, (d) $\lambda_c = 1.0$.

that the proposed method is suitable for performing PCCE in a wide range of $R$ values.

Furthermore, we performed a cross-validation test that measured the performance of trained deep learning models on other datasets to confirm the generalization ability of the ACE network. We conducted the cross-validation using the BSD [34] and LIVE [35] as training and test set, respectively. As shown in Fig. 8, without an additional training procedure, the proposed method exhibits superior performance as compared to the conventional methods. For instance, as shown in the first row of Fig. 8, [3], [8], and [11] degrade image details in the area of the wheel and wing of the airplane, *i.e.* yellow and red boxes, respectively. In contrast, the proposed method effectively preserves the details as well as enhances the contrast while reducing power consumption.

### B. Quantitative Comparison

The quantitative assessment of the PCCE algorithm is a challenging task, as there is no universally accepted quality assessment method for resultant images of PCCE. In this study, we adopt three different metrics for PCCE quality assessment: EME [36], normalized discrete entropy (NDE) [37], and VSI [12]. 1) The EME approximates the average contrast in an image by dividing the image into blocks, computing a score based on the minimum and the maximum gray-levels in each block, and averaging the scores. 2) The NDE measures the amount of information in an image. A higher NDE score means that the image contains richer details. 3) The VSI evaluates the image quality by comparing the perceptual similarities between the input and output images.

Table I and Table II provide the comprehensive performance benchmarks between the proposed and conventional methods. As shown in Table I, both the methods in [3] and [8] achieve high EME scores by enhancing the contrast. However, the HPCCE fails to preserve the perceptual similarity resulting in a low VSI score, whereas the method of Chang *et al.* [8] achieves a poor NDE score. Although the method of Chondro *et al.* [11] shows satisfactory performance in terms of VSI and NDE, this method exhibits an inferior EME score, because it mainly focuses on preserving image details without considering the contrast. These observations indicate that the conventional methods fail to improve the image contrast while maintaining the details and structural similarity of the image. The proposed method accomplishes not only a competitive EME score compared with conventional methods but also high NDE and VSI scores. Moreover, as shown in Table II, the proposed method exhibits superior performance than the LAPSE method [7]. These results demonstrate that the proposed method is suitable for performing PCCE without losing details and perceptual similarity.

To reveal the effectiveness of the $L_P$, in addition, we measured the actual power saving rate $R_a$ which indicates the $R$ of the $f(X_D)$. As depicted in Table I, the $R_a$ has the almost same value with the target $R$ value since we provided more constrained the $L_p$ compared to other loss terms; the ACE network can produce the output image with high quality while accomplishing $R$ that we selected.

### C. Ablation study

In the proposed method, the loss function consists of three major terms: *power*, *similarity*, and *contrast loss*. To show the effectiveness of each loss function, we trained the ACE network without each loss function. Fig. 9 shows the effectiveness of each loss term. As depicted in Fig. 9(b), the ACE network trained without the *power loss* produces the high contrast image without reducing power consumption. Thus, this approach cannot constraint the power of OLED display. As shown in Fig. 9(c), the ACE network trained without the *similarity loss* fails to preserve the image contents. (The color in-formation is preserved since the ACE network only handles the luminance of input image.) The effectiveness of the *contrast loss* is shown in Fig. 9(d). Compared with the resultant image of the proposed method, which is depicted in Fig. 9(e), the ACE network trained without the usage of the *contrast loss* fails to enhance the contrast. These results reveal that three loss terms jointly work to produce the high contrast images while preserving the image contents as well as constraining the power consumption.

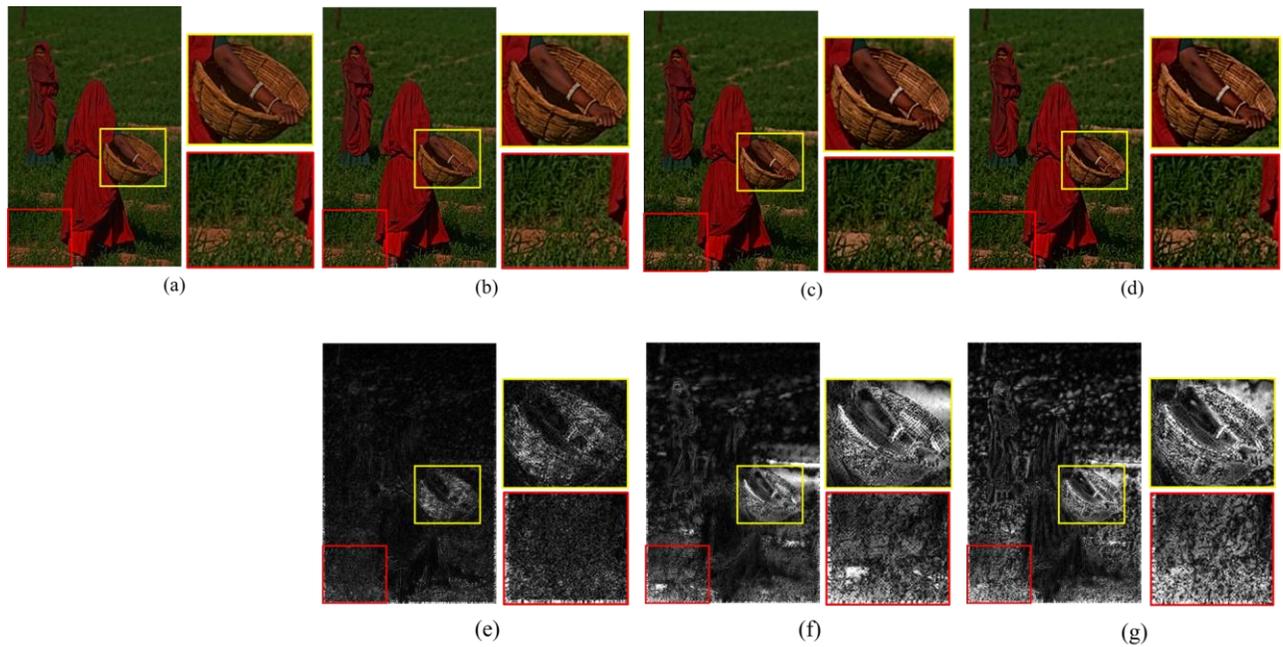

Fig. 11. The visual comparison of the proposed method trained by using different loss functions: (a) without contrast loss, (b) with local contrast loss, (c) with global contrast loss, and (d) with local and global contrast losses. (e)-(g) indicate the corresponding difference maps between (a) and (b)-(d), respectively. In our experiments, we set $R$ to 0.7.

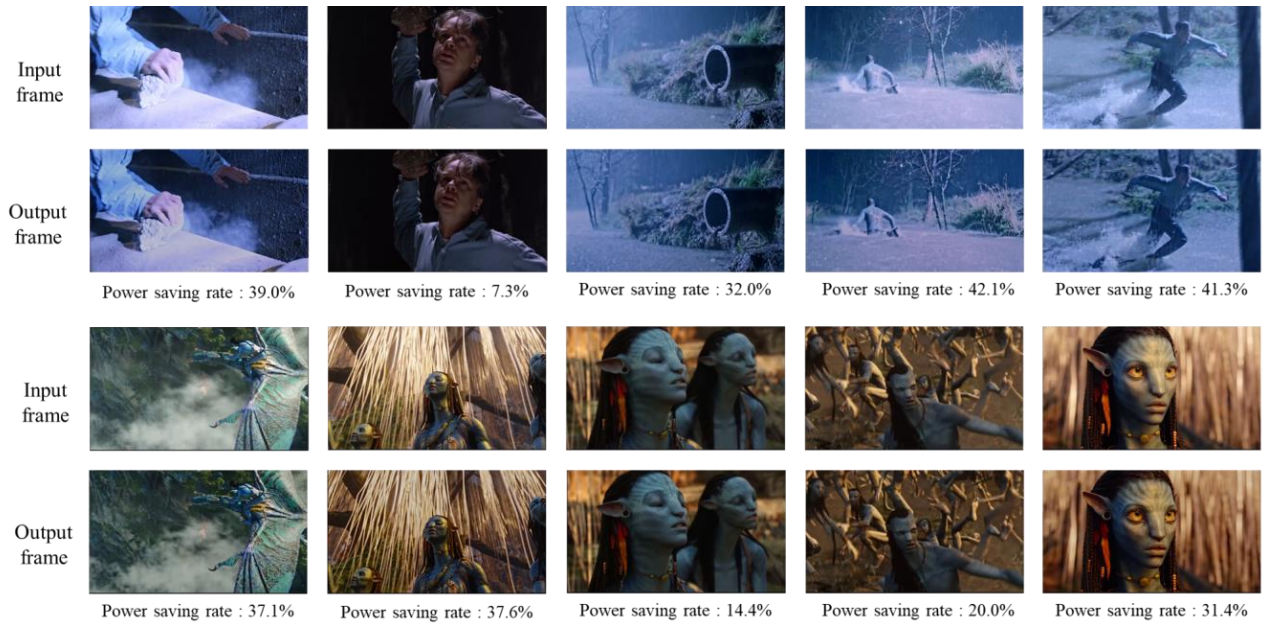

Fig. 12. The selected frame of video clips. The proposed method not only generates the visually pleasing output image but also effectively constraints the power.

To investigate how $L_c$ influences on enhancing the contrast of the resultant images, we conducted experiments by varying the values of $\lambda_c$. Fig. 10 depicts the resultant images on the BSD test set. As shown in Fig. 10(a) and Fig. 10(b), the ACE network with large $\lambda_c$ value (=0.25) more enhances the contrast compared to small $\lambda_c$ value (=0.1). These results indicate that the degree of the contrast enhancement can be determined by altering the $\lambda_c$ value.

However, as shown in Figs. 10(c) and 10(d), a large $\lambda_c$ may result in unexpected distortions owing to over-enhancement. These problems are caused by the *similarity loss* and *contrast*

*loss* which compete with each other. More specifically, the *similarity loss* encourages the ACE network to generate the resultant image similar to the input image. In contrast, the *contrast loss* guides the ACE network to maximize the local and global variances irrespective of the preservation of image contents. Thus, for larger $\lambda_c$, the *contrast loss* term surpasses the *similarity loss*, making the ACE network produce the image with a high variance regardless of the preservation of image contents. To control the importance of two loss terms, *i.e. similarity loss* and *contrast loss*, we empirically set $\lambda_s$ and $\lambda_c$ as 2 and 0.25, respectively.

On the other hand, as mentioned in Section III.B, the *contrast loss* is the combination of two different loss terms, *i.e.*, global and local contrast loss terms. Thus, we conducted ablation studies which investigate the effect of each loss term. In our experiments, we trained the ACE network using four different loss functions: (*i*) power loss + similarity loss, *i.e.* ACE network trained without *contrast loss*, (*ii*) power loss + similarity loss + local contrast loss, (*iii*) power loss + similarity loss + global contrast loss, and (*iv*) power loss + similarity loss + global contrast loss + local contrast loss. In the following subsection, we use (*i*)–(*iv*) to refer to these loss functions. As listed in Table III, (*i*) shows the lowest performance in terms of EME, whereas (*iv*) exhibits the highest performance. In addition, Fig. 11 shows the qualitative results of the aforementioned loss functions. In order to effectively show the role of each loss term, as shown in Figs. 11(e)-(g), we compute the difference map between the output image of (*i*) and those of other loss functions, *i.e.* (*ii*)-(*iv*). As shown in Fig. 11(b) and Fig. 11(e), the local contrast loss enhances the local details such as basket region. The global contrast loss improves the overall image contrast as illustrated in Fig. 11(c) and Fig. 11(f). By combining the local and global contrast losses, the (*iv*) effectively enhances the contrast of whole image including texture regions such as grass area. These results demonstrate that our *contrast loss* effectively guides the ACE network to enhance the contrast.

### D. Performance Evaluation in Video

To evaluate the performance of the proposed method in video sequences, we conducted experiments for two different video clips [3]: "The Shawshank Redemption" and "Avatar." Motivated by [3], we employ the average luminance of the input video frame as the *R* value of ACE network, which is expressed as follows:

$$R_i = \overline{Y}^\rho, \qquad (9)$$

where $\overline{Y}$, $R_i$, and $\rho$ indicate the average luminance in the range [0, 1], the *R* value at the *i*-th frame, and a power-control parameter, respectively. We set $\rho$ to 1.5 in our experiments. For bright input frames with high $\overline{Y}$, $R_i$ is set to a large value to achieve aggressive power saving, whereas for dark input frames, it is set to be close to 0 to avoid the brightness reduction. Fig. 12 represents the randomly selected frames. As illustrated in Fig. 12, the proposed method effectively saves power while providing visually pleasing image quality. The computational time of the proposed method was 31.2 ms on average with 1280 × 720 image resolution, which indicates that real-time performance is guaranteed.

## V. CONCLUSION

In this work, we have introduced a novel unsupervised PCCE model for the OLED display. We demonstrated that the PCCE technique can be performed using a simple and compact CNN called the ACE network. To train the network without the reference image, we proposed a novel loss function which guides the network to enhance the contrast while preserving the structural information and constraining the power consumption. We provided quantitative and qualitative comparisons on BSD and LIVE dataset. Experimental results demonstrated that the proposed method not only produces visually pleasing image but also achieves promising performance in terms of EME, NDE, and VSI. It is expected that the proposed method, with the advantage of simplicity and robustness, is applicable to other low-level vision problems such as image deblurring and low-light image enhancement.

This paper is the first work on using the unsupervised learning scheme to PCCE successfully. However, the proposed ACE network requires exhaustive computational power owing to the usage of the GPU. This problem can be addressed in the OLED display by embedding the pre-trained network parameters to the FPGA-based CNN accelerators which have high computational and power efficiency. As our future work, we will further improve the proposed ACE network by reducing the computational complexity and power.

## BIOGRAPHIES

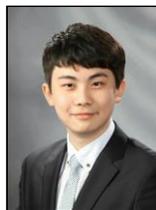

**Yong-Goo Shin** received his B.S. degree in Electrical Engineering from Korea University in 2014. He is currently pursuing his Ph.D. degree in Electrical Engineering at Korea University. His research interests are in the areas of human–computer interface, digital signal processing, computer vision, and artificial intelligence.

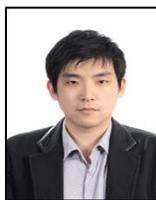

**Seung Park** received his B.S. degree in Electrical Engineering from Korea University, Seoul, Korea, in 2013. He entered the Computer Vision and Image Processing Lab in the Department of Electrical Engineering of Korea University in March 2013, and is currently pursuing a Ph.D. degree. His research interests include image processing, contrast enhancement, visual tracking, and visual odometry.

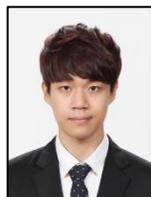

**Yoon-Jae Yeo** received his B.S. degree in Electrical Engineering from Korea University in 2017. He is currently pursuing his Ph.D. degree in Electrical Engineering at Korea University. His research interests are in the areas of image processing, computer vision, and deep learning.

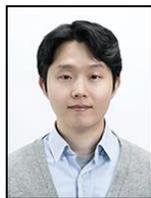

**Min-Jae Yoo** received the M.S. Degrees in Electrical Engineering from Korea University, Seoul, Korea, in 2018. He is working as a research engineer in LG Display. His research interests include image processing and computer vision.

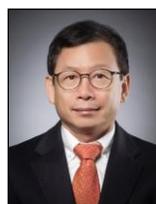

**Sung-Jea Ko** Sung-Jea Ko (M'88-SM'97-F'12) received his Ph.D. degree in 1988 and his M.S. degree in 1986, both in Electrical and Computer Engineering, from State University of New York at Buffalo, and his B.S. degree in Electronic Engineering at Korea University in 1980.

In 1992, he joined the Department of Electronic Engineering at Korea University where he is currently a Professor. From 1988 to 1992, he was an Assistant Professor in the Department of Electrical and Computer Engineering at the University of Michigan-Dearborn. He has published over 180 international journal articles. He also holds over 60 registered patents in fields such as video signal processing, computer vision, and multimedia communications.

Prof. Ko is the 1999 Recipient of the LG Research Award. He received the Hae-Dong best paper award from the Institute of Electronics and Information Engineers (IEIE) (1997), the best paper award from the IEEE Asia Pacific Conference on Circuits and Systems (1996), a research excellence award from Korea University (2004), and a technical achievement award from the IEEE Consumer Electronics (CE) Society (2012). He received a 15-year service award from the TPC of ICCE in 2014 and the Chester Sall award from the IEEE CE Society in 2017. He has served as the General Chairman of ITC-CSCC 2012 and the General Chairman of IEICE 2013. He is a member of the editorial board of the IEEE Transactions on Consumer Electronics. He is a distinguished lecturer of the IEEE. He was the President of the IEIE in 2013 and the Vice-President of the IEEE CE Society from 2013 to 2016. He is a Fellow of the IEEE (2012) and a Fellow of the Institution of Engineering and Technology (IET) (2000).